\documentclass[letterpaper,10pt]{article}
\usepackage[spanish]{babel}
\usepackage{longtable}
\usepackage[latin1]{inputenc}
\usepackage{multicol}
\usepackage[dvips]{graphicx}
\usepackage{subfigure}

\title{\huge{\textbf{Sobre la correlación entre las micro explosiones en el radio y el cambio de los parámetros del viento solar}}}
\author{Juan Carlos Martínez Oliveros\\ Daniel Ricardo Izquierdo Peña}
\begin{document} 
\maketitle
%\selectlanguage{spanish}
\begin{abstract}
El Sol es la estrella más cercana a nuestro planeta y por ello a sido la más estudiada en el transcurso de los años, sin embargo, aún es mucho lo que desconocemos de ella y aún más lo que no hemos podido comprender. Uno de los fénomenos solares que tiene repercución directa con la Tierra es el viento solar. El viento solar se define como el plasma que es expulsado desde las capas superiores de la atmósfera solar, este viento ha sido catalogado y en la actualidad se considera que posee tres componentes. 

\begin{enumerate}
\item El viento solar pasivo: Es la componente del viento solar que existe como un flujo constante de plasma, y que llena el espacio interplanetario.
\item Flujo supersónico y cuasiestacionario.
\item Flujos supresónicos esporádicos. Estos son debidos a procesos cortos duración en el Sol, que son isotrópicos y además muy complejos en su estructura.
\end{enumerate}

Se presenta una explicación breve del modélo de Parker del viento solar y el análisis de la correlacion entre radio explosiones en el Sol y el cambio en los parametros del viento solar. 
\end{abstract}

\renewcommand{\abstractname}{Abstract}
\begin{abstract}
The Sun is the closest star to our planet and it is the most studied, perhaps,  exist too much procesess not-understood. One of the solar processes that have a direct interaction with the earth is the solar wind. The solar wind is defined as the plasma expulsed from the solar atmosphere, this wind was cataloged  and is considered that have three components:

\begin{enumerate}
\item Passive solar wind: Is the constant component of the solar wind.
\item Supersonic and quasistady flux.
\item Sporadic supersonic flux.
\end{enumerate} 

We present and brief explanation of the Parker's model of the solar wind and a correlation analysis between solar micro radio bursts and the change of the solar wind parameters.

\end{abstract}

%\begin{multicols}{2}
\section{El viento solar}

El viento solar es considerado al igual que el campo magnético interplanetario (IMF) una extensión de la corona solar. Existen modelos teóricos que descríben las variaciones de los parámetros termodinamicos  (densidad, presion, temperatura, volumen) en funcion de la distancia al Sol, es decir, el radio solar. Chapman en los años 50 propuso el primer modelo, en su análisis Chapman describe la corona como un medio homogeneo y simétrico lo cual le permitió hacer una caracterizacion del medio usando la hidrostática.

Parker \cite{parker1} propuso un modelo din\ amico de la corona, en el que se analiza la probabilidad de flujo variable de partículas de la corona. Estas ideas fueron basadas en las observaciones realizadas por Biermann\cite{biermann}. En sus observaciones Biermann registro que la longitud aparente de la cola de los cometas variaba con el tiempo y estas se hacian mucho mas evidentes en los momentos de mayor actividad solar. Biermann propuso que esta variacion es debida a un flujo variable de particulas del Sol. 

\subsection{El modelo de Parker}

El modelo de Parker nos describe la atmósfera solar desde el punto de vista de la magnetohidrodinámica. La atmósfera solar se dice que se encuentra en equilibrio y que el flujo de partículas (electones, protones y partículas alfa), por simetría, se espera que sea radial. La ecuación de conservación del momento radial para la corona es,

\begin{center}
\begin{equation}
\rho u \frac{du}{dr} = -\frac{dp}{dr} - \rho \frac{GM_o}{r^2}
\end{equation}
\end{center}

donde $u$ es la velocidad radial de expansión. La ecuación de continuidad es

\begin{center}
\begin{equation}
\frac{1}{r^2} \frac{d(r^2\rho u)}{dr} = 0
\end{equation}
\end{center}

esta también puede ser escrita en función del área transversal por la cual fluye el plasma

\begin{center}
\begin{equation}
\rho(r)u(r)A(r) = \rho_0 u_0 A_0
\end{equation}
\end{center}

La ecuación de estado del gas esta dada por una función politrópica  que relaciona la presión y la densidad del medio,

\begin{center}
\begin{equation}
p(r) = p_0 \left( \frac{\rho(r)}{\rho_0} \right) ^{\alpha}
\end{equation}
\end{center}

donde $\alpha$ es el índice politrópico y entre$1 \le \alpha \le 5/3$ y corresponde al coeficiente adiabático para $\alpha = 5/3$. Indices distintos a 5/3, implican la existencia de otras fuentes energéticas del viento solar. Reemplazando los valores de la ecuación de continuidad y de estado del gas en la ecuación de conservación del momento e intergrado por $r$ , obtenemos la ecuación de Bernulli para $\alpha \neq 1$

\begin{center}
\begin{equation}
\frac{1}{2}u^2 - \frac{GM_o}{r} + \frac{\alpha}{\alpha -1}\frac{p_0}{\rho_0}\frac{u_0A_0}{uA} = \frac{1}{2}u_0^2 - \frac{GM_o}{r_0} + \frac{\alpha}{\alpha -1}\frac{p_0}{\rho_0}
\end{equation}
\end{center}
 
reescribiendo esta ecuación en coordenadas adimensionales,\

\begin{eqnarray}
\zeta = \frac{r}{r_0}, & v^2 = \frac{1}{2} \frac{p_0}{r_0} u^2, & H = \frac{GM_o\rho_0}{r_0p}, \nonumber
\end{eqnarray}

\begin{equation}
v^2 + \frac{\alpha}{\alpha -1}\left( \frac{v_0}{v \zeta^2}\right)^{\alpha -1} -\frac{H}{\zeta}  = v_0^2 + \frac{\alpha}{\alpha -1} - H \equiv V_1^2
\end{equation}

donde $V_1$ es una constante de integración que depende de las condiciones de frontera existentes sobre la superficie $r = r_0$.

Esta ecuación nos muestra la variación de la velocidad del viento solar con respecto a la distancia al Sol. Esta ecuación no posee solución analítica, sin embargo han sido calculadas en diversas soluciónes en distintos trabajos. Se mostrarán a continuación las soluciónes asintóticas para distancia grandes ($\zeta \gg 1$) y pequeñas ($\zeta \ll 1$).

\subsection{Soluciones asintóticas}

\subsubsection{$\zeta \rightarrow \infty$}

Podemos observar que si $\zeta$ tiende a infinito, entonces la función $v(\zeta)$ puede crecer infinitamente, tender a cero o a una constante. Podemos observar de la ecuación anterior que $v(\zeta)$ no puede tender a infinito, ya que el término de la derecha es una constante. Así nos queda que la función $v(\zeta)$ tiende o a una constante o al cero. Supongamos que $v(\zeta)$ tiende a una constante,

\begin{equation}
v\vert_{\zeta \rightarrow \infty} \longrightarrow v_1
\end{equation}

Si $v(\zeta)$ tiende a cero, el primer y tercer elementos del lado izquierdo de la ecuación  también tiende a cero, así,

\begin{equation}
v\vert_{\zeta \rightarrow \infty} \longrightarrow \frac{v_0}{\zeta^2}\left( \frac{\alpha}{\alpha -1}\frac{1}{v^2}\right)^{\frac{1}{\alpha -1}} 
\end{equation}

De esta forma podemos observar que este caso posee dos soluciones, ahora debemos determinar cual de ellas es más aproximada a la realidad física, para ello calculamos la densidad del plasma, correspondientes a estas soluciones.

Tenemos que la ecuación de estado del gas en las nuevas coordenadas viene dada por,

\begin{equation}
\rho(r) = \rho_0 \frac{1}{\zeta^2}\frac{v_0}{v}
\end{equation} 

Reemplzando los valores obtenidos de $\zeta$ encontramos la densidad del plasma

\begin{equation}
v|_{\zeta \rightarrow \infty}  \cases{0 \cr \rho\left[ \frac{(\alpha -1)v_1^2}{\alpha}\right]^{\frac{1}{\alpha -1}} \cr}
\end{equation} 

De esta ecuación se deduce que la solución inferior de $v(\zeta)$ no puede tener lugar ya que para $v \rightarrow \infty$, el valor de la densidad tiende a una constante, lo cual es refutado por los resultados experimentales. La solución superior está de acuerdo con la experimentación ya que a grande distancias de Sol, la densidad de partículas debe tender a cero.

\subsubsection{$\zeta \rightarrow 0$}

Cuando en $\zeta \rightarrow 0$ el tercer término del lado izquierdo de la ecuación tiende a infinito, como el término de la derecha es una constante, entonces el crecimiento de este término debe de ser compensado por los dos términos del lado izquierdo de la ecuación, de esta forma tenemos de nuevo dos posibles soluciones.

\begin{equation}
v|_{\zeta \rightarrow 0}   \cases{\left(\frac{H}{\zeta}\right)^{1/2} \rightarrow \infty \cr v_0 \left[\frac{\alpha}{\alpha -1}\frac{1}{H}\right]^{\frac{1}{\alpha-1}} \zeta^{\frac{1}{\alpha-1}-2} \cr}
\end{equation} 

%\end{multicols}

\begin{figure}[ht]
\begin{center}
\includegraphics[scale=0.45]{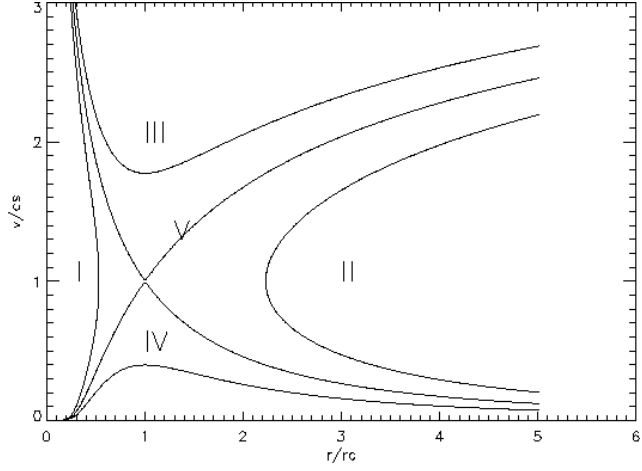}
\caption{En esta gráfica se puede observar el punto crítico de las soluciones.}
\end{center}
\end{figure}
%\begin{multicols}{2}

La solución superior no puede existir físicamente, ya que la velocidad del viento no puede crecer infinitamente a medida que nos alejamos del Sol. La solución inferior nos muestra un resultado aceptable cuando $v|_{\zeta \rightarrow 0}\rightarrow 0$, para los valores del índice politrópico definidos por la desigualdad $1/(\alpha - 1) -2 >0$, es decir, $\alpha < 3/2$.

De esta manera, la solución estacionaria de la corona se hace posible en el caso de que el índice de la politropa \textit{a} sea menor que el índice adiabático $\alpha = 5/3$, es decir, se tiene un flujo de energía constante en la corona y en el viento solar. En el modelo original de Parker se supone que este flujo es garantizado por la alta conductividad térmica del plasma, sin embargo, estudios posteriores demostraron que este flujo no es suficiente para garantizar la aceleracion del viento solar, y así es necesario el buscar nuevas fuentes energéticas.

Como ya pudimos ver, existen dos soluciones tanto para distancias grandes, como para cortas. La comparación de estas dos soluciones se hace observando su comportamiento cerca de un punto crítico, punto que en el plano $(\zeta,v)$, se determina de la siguiente forma.

Diferenciamos la ecuación de Bernoulli (en unidades adimensionales) con respecto a $\zeta$:

\begin{equation}
\left(2v - \frac{\alpha v_0^{\alpha-1}}{v^{\alpha}\zeta^{2(\alpha-1)}}\right)\frac{dv}{d\zeta} = \frac{2\alpha v_0^{\alpha-1}}{v^{\alpha-1}\zeta^{2(\alpha-1)+1}}-\frac{H}{\zeta^2}
\end{equation} 

Determinamos el punto crítico $(\zeta_c,v_c)$ como el punyo en el cual el lado derecho de la ecuación anterior y el coeficiente con $du/d\zeta$ se hacen iguales a cero al mismo tiempo. Entonces

\begin{eqnarray}
v_c^2\zeta_c=\frac{H}{4}, & \zeta_c = \left(\frac{H}{4}\right)^{\frac{\alpha+1}{5-3\alpha}}\left(\frac{2}{\alpha v_0^{\alpha -1}}\right)^{\frac{2}{5-3\alpha}}
\end{eqnarray} 

%\end{multicols}
%Insertar dos imagenes del Radio, donde se vea la micro, 21/01/2000
\begin{figure}[ht]
\linespread{0.55} \centering \subfigure[\footnotesize{Observación en tres frecuencias del radiotelecopio RATAN-600. Fecha de observación 21/01/2000  08:16 UTC. En esta imagen podemos observar un ejemplo de micro explosión.}]{\label{fig:subfig:a}
\includegraphics[scale=0.25]{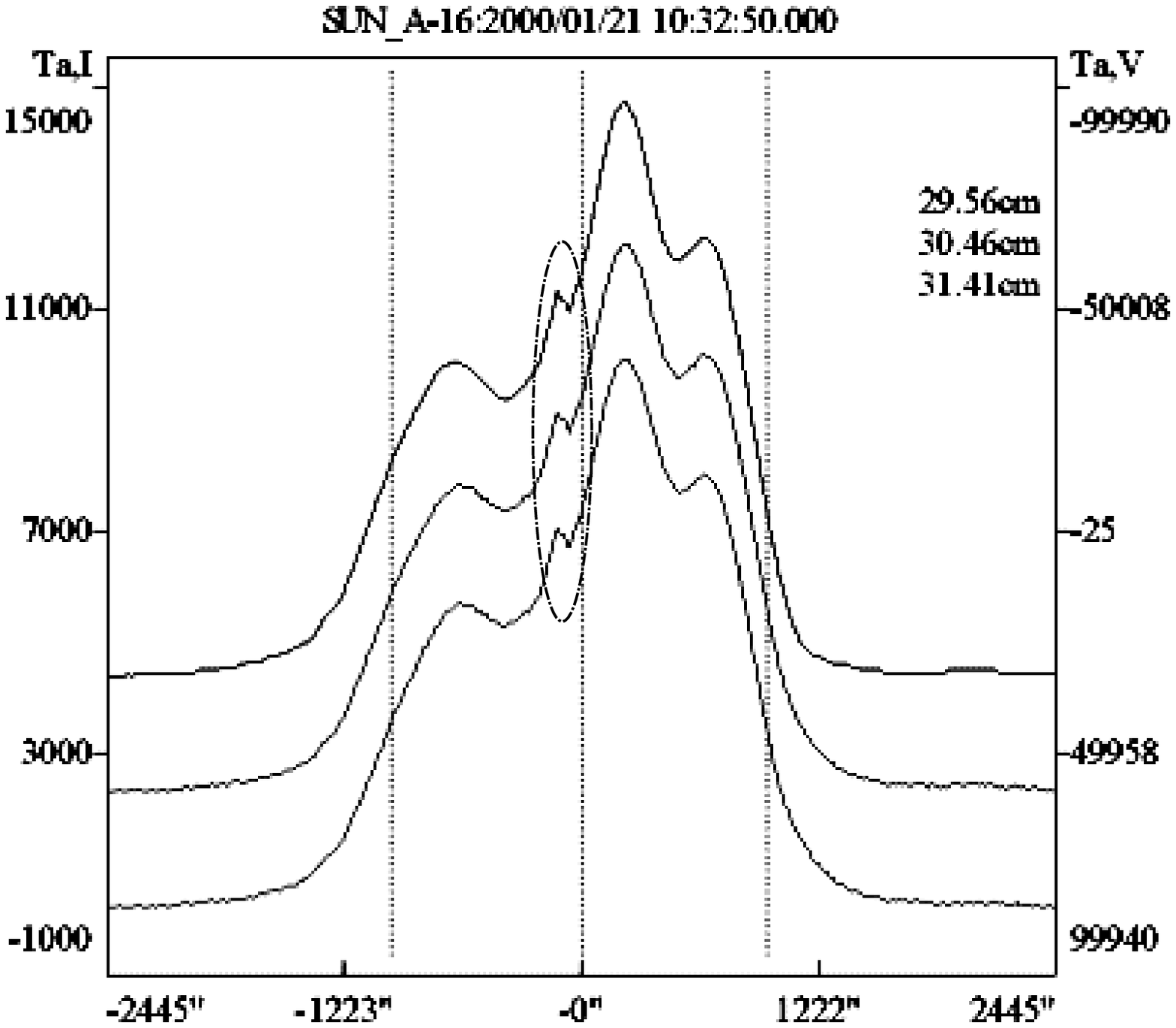}}
%%\hfill 
\subfigure[\footnotesize{Observación en tres frecuencias del radiotelecopio RATAN-600. Hora de observación 09:24 UTC.}]{\label{fig:subfig:b}
\includegraphics[scale=0.25]{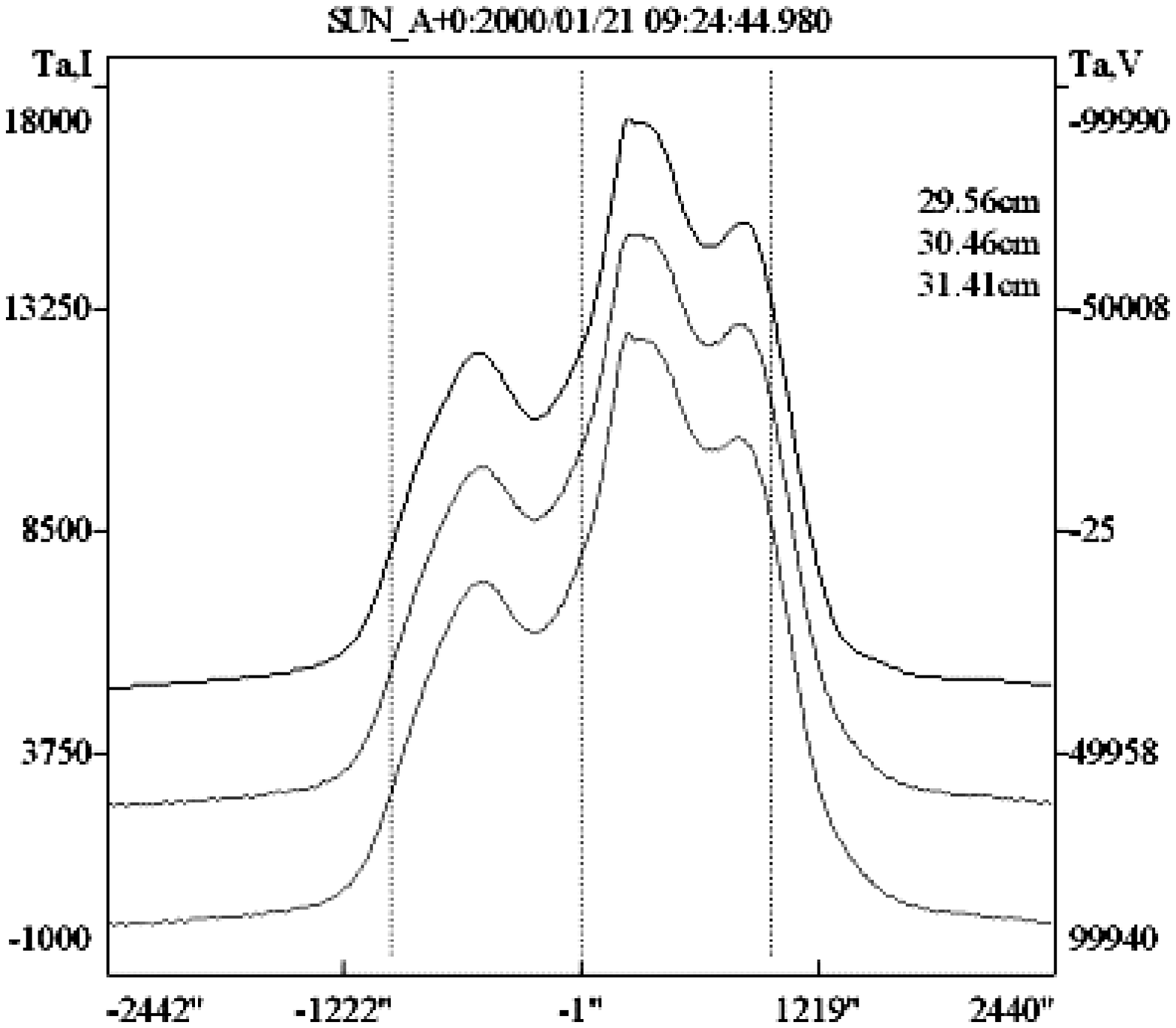}}
\caption{}
\label{fig:subfig}
\end{figure}
%\begin{multicols}{2}
El comportamiento de las soluciones cerca de este punto crítico se puede observar en la gráfica 1. Como podemos ver las soluciones son representadas como una familia de hipérbolas, sin embargo, sólo una de las soluciones safisface las condiciones de frontera tanto para grandes, como para pequeñas distancias al Sol. Como resultado se tiene que para distitas temperaturas la velocidad radial del viento solar varia entre 300km/s y 700km/s. Podriamos decir que esta variación en la velocidad es debida a la variación de la temperatura en la corona solar, lo cual es predicho por el modelo de Parker, sin embargo, distintas observaciones han comprobado que la fuente de flujo a grandes velocidades son los huecos coronales, en los cuales la temperatura de la corona es menor que la media. Así notamos que la velocidad del viento solar no sólo depende de la temperatura de la corona, sino también del índice politrópico $\alpha$: entre mayor será el valor de $\alpha$, la velocidad del viento solar sera mucho menor en la órbita terrestre. La mejor correlación entre modelos y observaciones para el modelo de Parker es $\alpha =1$ en la vecindad solar y $\alpha =5/3$ a grandes distancias del Sol.

Ahora, en relación con el valor inferior de $\alpha$, se tiene que el gradiente de temperatura tiende a cero $\partial T/ \partial r \rightarrow 0$. Además el flujo también tiende a cero. De esta forma, para poder mantener una temperatura suficientemente alta del viento solar, debe existir otro mecanismo energético no térmico, el cual puede estar relacionado con la energía disipativa de las ondas de Alfvèn.

%%%%%%%%%%%%%%%%%%%%%%%%%%%%
\subsection{El Campo Magnético Interplanetario}

El campo magnético del Sol ha sido medido de diferentes formas desde la Tierra y es consederado del orden del gauss. El principal mecanismo de generación del campo magn\ ético solar es el efecto de dínamo. Este campo general se extiende radialmente y decrece con el cuadrado de la distancia $r^{-2}$; se considera que el campo es de $\sim 2 \times 10^-5$ gauss \cite{space_aph} sobre la órbita terrestre. Es sobre estas líneas de campo que se pueden detectar las partículas cargadas que provienen de los eventos solares tales como protuberacias. El cambio en los parámetros del viento solar poseen una correlacion directa con los periodos de actividad solar \cite{asp}.

Este campo radial se transforma a distancias grandes del Sol en una espiral de arquimedes, debido a la rotacion de 27 días del Sol. De esta forma el campo en la órbita terrestre se vuelve más azimutal que radiasl, y decrece con $r^{-1}$.
La ``radialidad'' de las líneas depende de la velocidad del viento solar. Es sabido que esta velocidad varia según los procesos de generación del viento.  Las líneas de campo vienen dadas por:

\begin{equation}
r \cong [v(\infty)/\Omega]\phi
\end{equation} 

Donde $\Omega$ es la velocidad angular media del Sol y $\phi$ es el ángulo acimutal que es medido sobre la linea de campo que emerge sobre la superficie solar.

\section{Observaciones y Conclusiones}

Para el estudio de la correlación fueron escogidos 75 candidatos. Estos candidatos son micro explosiones detectadas en el radio diapazón y registradas por el radiotelescopio RATAN 600 de la Academia Rusa de Ciencias, en las frecuencias correspondientes a las longitudes de onda 29.56cm, 30.46cm y 31.41cm. Como micro explosión entenderemos los eventos de corta duración y que son vistos casi con la misma intensidad a distintas frecuencias y que no pueden ser asociados a problemas de amplificación del radiotelescopio (fig. 2).

Para este estudio, primero se halló la correlación con los eventos registrados por el satelite GOES en rayos X en las bandas de 0.1-8\AA y 0.1-4\AA. Se determinó que en el momento de cada microexplosión fue registrado con un error de $\pm5$ minutos, un incremento del flujo integral en rayos X en dos bandas, siendo este incremento más apreciable en la banda de 0.1-4\AA, sin embargo se tomó como patrón para el análisis de la correlación la banda de 0.1-8\AA, ya que esta es la usada en lo sumarios para la descripción de las explosiones y tormentas magnéticas.

%\end{multicols}

\begin{figure}[ht]
\begin{center}
\includegraphics[scale=0.30]{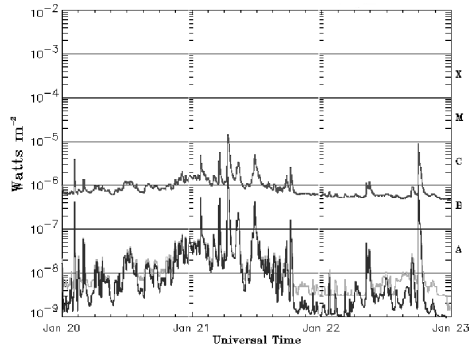}
\caption{Observación en rayos-x, hecha por el satelite ACE.}
\end{center}
\end{figure}
%\begin{multicols}{2}

Teniendo en cuenta que las micro explosiones no son eventos aislados, sino que hacen parte de procesos más energéticos, se implementó un método de acotación de los eventos tanto inferior como superiormente, para ello se analizaron los flujos en rayos X y se determinaron los eventos cercanos de mayor energía. Este método nos permite buscar dichos picos en las curvas de los parámetros del viento solar y ubicar por comparación de los flujo en el radio y del campo magnético del viento solar la micro estructura que fueron registradas en el radio diapazón. Se debe resaltar que este método podría dar luz sobre los posibles mecanismos de generación de la microestructura, esto a pesar de que las cotas que estamos tomando y los datos del viento solar sean integrales por el disco solar. 

Debido a que las líneas de campo magnético en el espacio interplanetrio están curvadas como es predicho por el modelo de Parker \cite{parker2}, se debe tener en cuenta que podemos no registrar todos lo eventos observados en el radio, es decir, ya que el viento solar se mueve a lo largo de las líneas de campo magnético los eventos ocurridos después de cierta longitud solar no puden ser registrados por los satélites de observación solar tales como el usado para este trabajo. Después de determinar aquellos eventos que pueden ser observados por el satélite ACE, fueron trabajados los 66 candidatos, determinando el tiempo en el cual fue registrado por el satélite ACE.

El análisis de la correlación solamente fue hecha suponiendo la pertenecia de la micro explosión a alguno de los dos eventos energéticos que la acotan, así fue posible realizar la tabla de correlación. Se establecio que la gran mayoria de las explosiones correspondientes en rayos-x a los eventos en el radio no son altamente energéticos y corresponden a clases B-C, siendo predominante las de clase C. 

Se obtuvó que el periodo de llegada de las particulas esta entre 3 y 6 días , siendo la principal componente los eventos solares de clase C con una occurencia del 67.09\%. En la siguiente tabla se muestran los resultados porcentuales:

\begin{center}

\begin{tabular}{|c|c|c|c|c|}
\hline
día & $<$C & C & M & V(km/s) \\ \hline
3 & 34.62\% & 57.69\% & 7.69\% & 577.16 \\
4 & 18.75\% & 68.75\% & 12.50\% & 482.87 \\
5 & 47.06\% & 50.00\% & 2.94\% & 346.26 \\
6 & 16.67\% & 66.67\% & 16.67\% & 288.58\\
\hline
\hline
\end{tabular} 
\end{center}

\begin{figure}[ht]
\begin{center}
\includegraphics[scale=0.50]{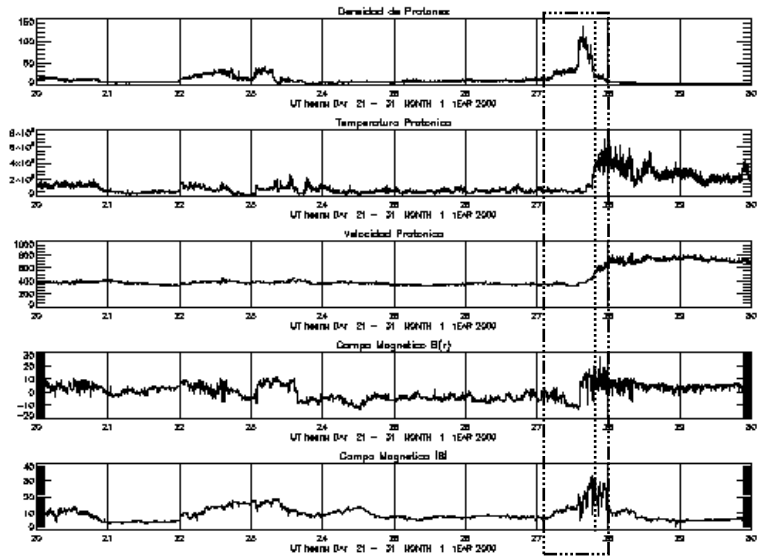}
\caption{Parámetros del viento solar.}
\label{sw}
\end{center}
\end{figure}

Todos los parámetros tomados ($\rho$, $T$ , $\vec{v}$, $|\vec{B}|$) para este trabajo presentan un cambio apreciable en su magnitud, sin embargo, los cambios mas notorios fueron en densidad de protones y temperatura. Fueron registrados un corrimiento de los máximos de los parámetros, estando el incremento de la temperatura y de la velocidad en un desfase con la densidad y el campo magnético (fig. \ref{sw}). Se observa una buena correlación entre el incremento de los flujos del viento solar tanto en densidad como en campo magnetico, demostrando una relación entre estos. En la gran mayoría de los casos los máximos coinciden entre estos parámetros y aún más los perfiles delos flujos coinciden muy bien. En cuento a la velocidad y la temperatura, el incremento de la velocidad para ciertos días no es muy apreciable, sin embargo se nota que la tendencia de aumento del flujo coincide con el incremento de la temperatura, y para los casos en los cuales el cambio de la velocidad es apreciable tambien se obtiene que los máximos de estos dos parámetros coinciden. Así se puede observar, al igual que en el caso de la densidad y el campo magnético, una relación directa entre estos dos parámetros, lo cual es claro de la teoría.

\renewcommand{\refname}{Referencias Electrónicas}

%%\end{section}
%\end{multicols}
\end{document}